\documentclass{article}
\usepackage{amsfonts,amssymb,amsmath,graphicx,graphics,marvosym,fullpage}
\begin{document}

\title{Open Access, Intellectual Property, and How Biotechnology 
Becomes a New Software Science}

\author{Fionn Murtagh \\
Head, Information, Communications \& Emergent 
Technologies Directorate \\
Science Foundation Ireland, Wilton Park House, Wilton Place, Dublin 2, Ireland \\
and \\
Department of Computer Science, Royal Holloway, University of London \\
Egham TW20 0EX, UK \\
Email fmurtagh@acm.org
}

\maketitle

\begin{abstract}
Innovation is slowing greatly in the pharmaceutical sector.  
It is considered here how 
part of the problem is due to overly limiting intellectual property 
relations 
in the sector.  On the other hand, computing and software in 
particular are characterized 
by great richness of intellectual property frameworks.  Could 
the intellectual property ecosystem of computing come to the aid
of the biosciences and life sciences?  We look at how the answer
might well be yes, by looking at (i) the extent to which a drug 
mirrors a software program, and (ii) what is to be gleaned from 
trends in research publishing in the life and biosciences.
\end{abstract}

\section{Introduction: Convergence of Computing and the Life Sciences}

The term ICT, information and communications technologies, will be 
used here to denote computing-in-the-large, including telecoms,
networking, hardware including photonics, design, applications, and 
so on.  We will use the term computing to be somewhat more restrictive,
involving software, software engineering, applications -- all 
areas built on computational thinking \cite{compthink}.  

ICT has a crucial role to play in drug creation; in early warning
and rapid response systems when faced with pandemics; in developing 
new, responsive technologies to face the threats of epidemics and 
pandemics.  But ICT goes way further than just an infrastructure 
for the life sciences or for any other area.  In fact \cite{kelty}, 
``Software and networks can express ideas in the conventional written
sense as well as create (express) infrastructures that allow ideas to 
circulate in novel and unexpected ways.''

\section{Intellectual Property in the Life Sciences: Current Focus on Patents}

\subsection{Crisis of Innovation in the Life Sciences}

Life sciences dominate the patent system.  
``Life sciences contribute the lion's share of patent revenues at 
leading US universities, outpacing contributions from physical science, 
information technology, and other fields. Life scientists also supply 
most of the inventions patented by the 10 technologically strongest US 
institutions''  according to \cite{agres}.
``In 2001, the top 10 US universities 
generated 689 life science patents, compared with 263 in information 
technology and 245 in all other technology categories. 

These life science discoveries contribute 42\% to the overall tech strength 
portfolio at these universities, compared with a 35\% contribution by 
information technology and a 23\% contribution from all other technology 
categories. The technical strength of a university indicates the extent 
to which its inventions influence others in their fields, as determined 
by how often its patents are cited by other patents. 

Life sciences contribute more than do other fields to institutions' 
technological and research strengths because ``companies in the life 
sciences have a lot of experience in commercializing ideas from 
universities,'' \cite{agres}, who continues:  
``[Companies in] engineering and 
other fields do not have this same experience because they tend to 
develop ideas themselves. The engineering fields also are more process 
oriented, while the life sciences are more product oriented. That makes a 
difference.''

Linking the patent system with innovation in universities, a 
journalist view \cite{raedupree} is: 
``it was the life sciences -- in particular, 
biotechnology -- that started universities down 
the slippery commercial slope in the first place. Even before the 
Bayh-Dole Act, pharmaceutical 
companies were eagerly trolling campuses, looking for projects to 
finance. After the law was 
passed, they stepped up their efforts, but now with renewed zeal 
for keeping potential trade 
secrets from competitors.'' 

There is however a growing problem with the 
innovation process in pharmaceuticals.  A leading researcher in 
the field has sketched out this situation as follows
\cite{fitzgerald}:
``The number of new drugs approved by the FDA has fallen linearly from 
53 in 1996 to 17 in 2007, the same number as in 1983. A slight bump 
upwards to 21 approvals in 2008 includes 3 drugs eventually approved 
on reconsideration and 3 radiocontrast agents. In other words, it 
doesn't buck the trend. This coincides with downward pressure on 
drug pricing. The growth in prescription drug sales -- 10\% of the 
\$286.5 billion US healthcare budget in 2007 -- had plummeted even 
before the present crisis: generics now account for roughly 60\% of 
the market and are rapidly growing in market share.
 
Pharma has reacted by shedding jobs in the US -- more than 100,000 
over the past 5 years -- and moving research to join drug production 
in lower cost economies overseas. Anticipated future revenue for the 
industry has shifted dramatically towards Asia. The current crisis is 
likely to accelerate these trends.''

Walton and Frank \cite{walton} give indicative figures of 80\% of
all scientists who ever lived being alive today; only 10--20 new
drugs approved each year; to get a drug to market it costs now
\$1 billion and takes 15 years.  Now, they state,
an estimated \$13.1 billion of drugs are going generic in 2008
and \$6.7 billion in 2009.

\subsection{Exclusivity of Exploitation by Pharmaceutical 
Companies and Its Loss}
\label{sect22}

Companies in the ``highly regulated and 
R\&D driven'' 
pharmaceutical sector are considered as (i) originating, driving research 
and the regulatory process including clinical trials, and protecting 
products through time-limited patent protection; and (ii) generic companies,
entering markets post-patent protection with equivalent products. 
Interestingly, 35\% of the compounds taken on board by originator companies
are acquired or licensed from third party, often small or medium sized 
biotech, companies.   

``The pharmaceutical sector is one of the main users of the patent system.''  
This is quite an interesting assertion, because it is so clearcut,
from the Executive Summary 
of the Final Report on the EC (European Commission) 
Competition Inquiry \cite{ECpharmafinal}.  
And: ``Contrary to what might be assumed, blockbuster medicines' patent portfolios
show a steady rise in patent applications throughout the life cycle of a product.  
Occasionally they show an even steeper increase at the end of the protection 
period conferred by the first patent'' 
\cite{ECpharmaprelim}.  

The EC initiated the inquiry into competitiveness in the pharmaceutical 
sector in January 2008, leading to 
the Final Report in 
July 2009 \cite{ECpharmafinal}.  Motivations were ``delays in the
entry of generic medicines to the market and the apparent decline in 
innovation as measured by the number of new medicines coming to the 
market''.  
Prescription medicines for human use only were at issue.  

The decline in novel medicines reaching the market was such that there
were on average 40 new medicines per annum between
1995 and 1999, and in the 2000s this has averaged 27 \cite{ECpharmafaq}.  
The effective patent protection period from product launch to the first
generic launch is over 14 years in 2009, up from 10 years 
in 2000 \cite{ECpharmafaq}. 
After about two years, generic companies have the effect of taking the 
erstwhile price for the drug down by 40\%.  
Generic market share varies a lot between countries: in Poland it is highest
at 56\%, whereas it is lowest in Ireland (13\%), France (15\%) and Finland (16\%)
\cite{ECpharmaprelim}.  

The industry distinguishes between
primary and secondary patenting.  The latter include ``different dosage forms, the production
process or ... particular pharmaceutical formulations'' \cite{ECpharmafinal}. 
Further, in order to prolong the exploitation of (time limited) protected pharmaceutical 
products use is made of numerous defensive patents, referred to as ``patent clusters'' or 
``patent thickets''.  
The Final Report notes that ``individual medicines are protected by up to nearly 100 
product-specific patent families, which can lead to up to 1,300 patents and/or pending 
patent applications...''.  
The Final Report notes some ``awareness by patent holders that some of their patents
might not be strong'', and also that defensive patenting has the clear benefit of 
slowing down the ability of generic products to come forward and take over the role of the
originating company's product.  

Outcomes of the Final Report are recommendations for a
faster and more efficient 
intellectual property system, including a (European) Community patent and a unified 
litigation system.  
Promised for 2010 is a new consideration ``on the use of personalized 
medicines 
and `-omics' technologies in pharmaceutical research and development 
and on the possible
need for new ... instruments to support them''.  

At issue here are ``new technologies like pharmacogenomics and 
patient-specific modelling and disease simulators'' for personalized 
medicine.  
The prospects for personalization of health and medical care based
on the association of human illness with genetic make-up is not always 
viewed with optimism though \cite{nyt16apr2009}.  This is due to the
complexity of common diseases like cancer and diabetes, linked to multiple
genetic variations.  

\section{Intellectual Property: Between Patents and Other Forms of 
Property Rights that Include Research Publishing}

\subsection{Implications of Pandemics for Patent Protection}

Apart from legal exclusivity running out, there is a further possible
stress point on exclusivity.  

To tackle the expected influenza A/H1N1 pandemic, in May 2009 
the World Health Organization
(WHO) allowed the leading Indian pharmaceutical corporate Cipla 
to produce the generic version of the anti-viral medication, Tamiflu.  
Hoffmann-La Roche's oseltamivir, branded as Tamiflu, and 
GlaxoSmithKline's (GSK) Relenza are the only two recognised drugs to 
treat the pandemic swine flu, H1N1 \cite{cipla-aug2009}.
Cipla brought a generic 
drug, zanamivir, substituting for oseltamivir to market in 
August 2009. The inhalatory drug zanamivir was branded as Virenza 
in India.
By August 2009, in view of the less critical threat posed by the 
A/H1N1 virus, a withdrawal of zanamivir was planned by 
regulatory authorities in India.  The Hoffman-La Roche drug,
Tamiflu, only was to be used.  Meanwhile GSK's Relenza is a 
pre-1995 drug and lacks patent protection in India.  

What is of note here is the readiness of generic companies to 
step into the production and distribution breech, and how such a
situation unfolds in a possibly socially threatening 
context.  

\subsection{The Changing Culture of Publishing and Open Access}
\label{puboa}

The first scientific journals, as we recognize them,
 go back to the 17th century.
Major aspects of the research process have remained 
close enough to how things were done those few hundred years ago, 
including the research reporting process through publication.

Increasingly now, open access is gaining ground.  What this 
implies is that journal or conference proceedings articles ought to be
put in institutional or discipline repositories six months
(or immediately) following
publication.

It is interesting to probe further and see who or what is
leading the charge towards open access.  This
movement is led clearly enough by the medical and health sectors.
Firstly it is in these areas that there is potentially a
resonance with the marketplace for commercialization,
and with an expressed need for application and deployment.  
(See section \ref{sect41} below.)
Secondly, medical and healthcare research is one opening -- one
vantage point -- in regard to the large life sciences sector.
Thirdly and most of all, it is organizations like the National
Institutes of Health that have gone furthest, most quickly too,
in introducing open access policies.  Between April and May
2008, for example, NIH policy of mandated depositing of peer-reviewed
publications in
PubMed Central became finalized \cite{nih}.

\subsection{The Slow Move Towards Open Innovation Models in the 
Life and Biosciences}

The rights management involved in open access can be seen as 
one element of a more general open innovation approach to 
intellectual property use and dissemination. 

Open innovation is one among various collective activities that have
been pursued in the ICT arena in particular.  Examples noted by 
Broglia \cite{broglia} include: Linus Torvalds and the development 
of the Linux operating system; Richard Stallman who established 
the GNU, ``GNU's not Unix'', repository of all levels of software
from the early 1980s onwards; in conjunction with GNU, there has been
the GPL, Gnu Public License or copyleft, which allows open redistribution 
and modification of software code so long as others can maintain these
rights; the SourceForge repository of open source projects which according 
to \cite{broglia} in July 2008 had 180,000 registered projects and 
1.8 million users.  Broglia proceeds to IBM's use of Linux; the Red Hat 
distributor of Linux, set up by Bob Young; and Wikipedia.  Extrapolating
from these efforts, the point is made as to how collective, freely 
contributed effort and work -- the ``online volunteer'' -- is a 
new, remarkable and by now fully established model.  The results
are seen too in Second Life group practices, Amazon's recommender  
system, and content on FaceBook, YouTube, eBay, and so on.  
Both free and commercial models exist side by side and are at times
even intertwined. 

Vingron \cite{vingron} sees data as being a side of the life 
sciences that should be accommodated by open access.  
An element of change is seen in data: data sets are huge and so must 
be published (as ancillary to the paper) online.  Meanwhile, 
``There is a strong push for all data to be 
public (genome sequence, protein structure)''.

Richard Jefferson's BiOS, Biological Innovation for Open Society,
www.bios.net, 
goes all the way, seeking a fully open innovation environment for 
data in the biosciences.  

Open innovation can go hand in hand with more distributed and 
less centralized forms of development, and indeed this can even 
lead to more democratic organizational forms with fewer barriers
to entry.
While there is much to say in favor of open innovation, it is nonetheless
clear that centralized and proprietary control can also lead
to more reliable, more resilient, more robust and more recovery-enabled,
forms of development.  In the healthcare sector, software systems
have been on occasion developed centrally, based on large 
state contracts, and have failed disastrously.  For some, the 
blame lies in thoroughly inadequate requirements gathering.  Others
have seen the culprit as the refusal to access an open innovation
model, incorporating open source.  Johnson \cite{johnson} presents
a wide-ranging overview of these divergent views, based on case 
studies and experience with the US Veterans Health Administration (VHA).  
In passing we note that Johnson \cite{johnson}
includes examples of where centralized healthcare ICT systems have
been developed and implemented with recognized success.

\section{Pharmaceuticals as the New Software}
\label{pharma}

\subsection{Pharmaceutical, Software and Other Goods}
\label{sect41}

Convergence between the computing sciences and the life sciences is
ongoing but with a great deal of, as yet, unrealized targets
\cite{kovac}.  We will now sketch out a scenario with clear implications
for how the underpinning information-based sciences and engineering could
relate to the life sciences in a very new way.

Consider for one moment the greatest product of all time in terms of 
return on investment and
development.  It is the drug Lipitor (atorvastatin calcium), used 
for cholesterol
treatment, that cost Pfizer half
a million dollars to develop, and had a return of over 
13 billion dollars in 2007 
\cite{inquiry}.  
The patent on Lipitor runs out in June 2011.  
This is only one  drug among many where
intellectual property rights are likely to change greatly in the
next few years.  Generics will step into their place, and will be 
priced far less.

Now consider software and pharmaceuticals.  Both act on the environment.
As such they have determinate inputs and they process these inputs
in determinate  ways.  Assume that user interaction is part and parcel 
of the software.  So keyboard, visual, haptic, voice, sound,
printer, display, 
etc.\ are input and output interfaces to the software's ambient 
environment.  A 
drug has interaction too, in terms of influencing or discouraging
cell growth, modifying in a positive or negative way the 
ecosystem of the human or animal, tempering inflammation or other
secondary effects, and so on.  The former is often macro level, and 
physical.  The latter is often micro or nano level, biological and 
chemical.  The essential analogies remain: both systems are
carrying out interactions of a determinate and predictable kind on their 
environment.  

Now consider how both are characterized by, or contrasted by, the 
following.

\begin{itemize}
\item Potentially huge costs to produce, refine:  A drug takes, say,
15 years to produce.  Consider the software in the Airbus 380 or 
Boeing 777 as a counterpoint to this. 
\item Validating shares engineering and art, statistical science
and social sciences.  But in neither domain (software, pharma)
are the physical
or natural sciences really to the fore.  (Energy may yet change 
the picture here, and bring to the fore -- for example -- the 
thermodynamics of information: see \cite{karnani}.)
\item Once developed and validated, the cost of reproducing software or
a drug is effectively zero.
\end{itemize}

A drug can change its use and delivery, so the molecule or protein 
``platform'' can change function or role.  Cf.\ secondary patenting 
discussed in section \ref{sect22}.  So it is for software.  
Computer software was always 
non-trivial to demarcate as a product.  With reference to 
the US Copyright Act of 1976, and the US Code section 17 of 
1980, Kelty \cite{kelty} notes that:    
``During the 1980s, a series of court cases helped specify what counted
as software, including  source code, object code (binaries), screen
display and output, look and feel, and microcode and firmware.'' 

Adaptability and modifiability are crucial.  
For software, 
``It is a peculiar feature of copyright law that it needs to be 
updated regularly each time the {\em media} change ...''   We can 
consider ``gramophones, jukeboxes, cable TV, photocopiers, peer-to-peer
file-sharing programs''.  Further, 
``new questions arise: how much change constitutes a new work, and thus
demands a new copyright license?  If a licensee receives one copy of a
work, to which versions will he or she retain rights after changes?  ...
is the XML document equivalent to the viewable document ...?  Where does 
the `content' begin and the `software' end?''.  
Kelty speaks of ``denaturalization'' of the software product.

So it is analogously with a drug.  
Faced with declining innovation in the pharma sector, and following 
the model of software that reuse is not simply a technical issue, then 
the thought arises as to how to harness collaboration.  
Some communities (e.g.\ humanities) would be aghast at liberal creation
of derivative works.  Other communities (e.g.\ computing and 
engineering) revel in reuse.  Kelty (\cite{kelty}, p.\ 291) includes 
biology in the reveling category.  Maybe the life sciences will follow
suit, involving for example that the ``compulsory licensing of 
pharmaceuticals [could be] open to analysis'' according to the terms offered 
in \cite{kelty} (p.\ 304).  

In regard to software, Boyle \cite{boyle} notes how a creative open 
``commons'' in property rights contributed so much to spurring on 
the software industry.  He notes, with worry, how synthetic biology
``which shares aspects of both software (programming in genetic 
code) and genetic engineering'' is threatened by the lack of 
open models of intellectual property.  

In his concluding remarks on the evolution of biotech, with a particular
focus on IPR (intellectual property rights), 
Pisano \cite{pisano} stresses the need for 
organizational and institutional innovation in biotechnology, 
in order for it to overcome the slowness of productivity (i.e.,
rarity of new drug discovery and successfully bringing to market).
In order to prevent under-exploited ``islands of expertise'' in biotech
greater
openness and disclosure are needed.  Recall that both research
publishing, and patenting and licensing, are forms taken by such 
expertise and knowledge.  Pisano focuses on the latter areas of 
intellectual property -- patenting and licensing.  
Among other recommendations, he proposes
greater disclosure of clinical trial data much earlier in the development
process, and not just after approval. 
While such disclosure does happen through journal publishing
(cf.\ section \ref{puboa}), when 
this happens is at the discretion of the sponsoring drug company,
and will not necessarily include negative as well as positive findings.  
Pisano's objective is to overcome the logjam of innovation in the
biotechnology sector, and to facilitate investment decisions. 

\subsection{Conclusions}

As patents run
out, and as competitive pressure prizes open current intellectual
property holdings, 
all of the pharma and bio sectors will be affected.  
The metaphor of a supernova would be apt.
This is not just pointing to a massive
economic and technological change.  Rather, in analogy with, say,
a black hole resulting from a supernova, it is a new source
of tremendous dynamism  -- a melting together of pharma and software.

The implications of this would be profound.  It would lead to a very
different health
system, including all of insurance, medical and public health domains.
To have some feel
for the
direction of events, one  need only look carefully at computer science
and the ICT
sector.  Computer science and engineering are characterized by a
wealth of
``models'' both in research and development and in the very rich
ecosystem of intellectual
property.
The dream is one where
``Everything that can be delivered digitally will be, at a cost
approaching zero, through a bandwidth nearing infinity'' \cite{pixar},
and IPR is adapted to fit this world view.

Creating of new drugs would result from ultra high dimensional
search and discovery in truly massive, semantically rich data stores.
Drug development would need a new
pharma-oriented information search and fusion infrastructure
at its core.  Our health system would be even more integrally
based on the information infrastructure.
Fortuitously or presciently, the
most advanced driver of open access in research outputs is the
biomedical and health sector.  The ground is
already being prepared slowly, through
data provision, for the search
engines that will power the new pharma informatics and health
informatics.

To envisage barriers between disciplines coming down is not a bad prospect.
Computer science and engineering may be very different in one or
two decades from now.
An implication of such change is in research publishing. This is
because one
thing about the way we carry out our research and scholarly work is
that we, across all science disciplines,  have
become very influenced by the biosciences.  Take for example how
journal citation
rates are based on just two previous years.  Extreme recency has
come to count
greatly in citation practices, and a very small number of highly
profiled journals tower
over all others.  We know how different the scene is in computer 
science and engineering (cf.\ \cite{moed}).  Perhaps the future
of scholarly publishing across many disciplines is closer than we
think to today's
variety of publishing practices in computing, both science and
engineering, both
theory and application, and embracing both open source and commercial
rights and privileges.

\end{document}